\documentclass[final,english]{elsarticle}
\usepackage[T1]{fontenc}
\usepackage[latin9]{inputenc}
\usepackage{varioref}
\usepackage{units}
\usepackage{amsmath}
\usepackage{amssymb}
\usepackage{graphicx}
\usepackage{esint}
\usepackage{caption}
\usepackage{subcaption}

\pdfpageheight\paperheight
\pdfpagewidth\paperwidth

\journal{Elsevier}
\usepackage{upgreek}
\usepackage[bookmarks,bookmarksopen,bookmarksdepth=6]{hyperref}
\usepackage{cases}
\usepackage{url}
\usepackage{lineno}

\usepackage{a4wide}

\makeatother

\usepackage{babel}
\begin{document}

\title{The influence of edge effects on the determination of the doping profile of silicon pad diodes}

\author[]{M.~Hufschmidt}
\author[]{E.~Fretwurst}
\author[]{E.~Garutti}
\author[]{R.~Klanner \corref{cor1}}
\author[]{I.~Kopsalis}
\author[]{J.~Schwandt}

\cortext[cor1]{Corresponding author. Email address: Robert.Klanner@desy.de,
 Tel.: +49 40 8998 2558}
\address{ Institute for Experimental Physics, University of Hamburg,
 \\Luruper Chaussee 147, D\,22761, Hamburg, Germany.}



\begin{abstract}

  Edge effects for square $p^+n$ pad diodes with guard rings, fabricated on high-ohmic silicon, are investigated.
  Using capacitance-voltage measurements of two pad diodes with different areas, the planar and the edge contributions to the diode capacitance are determined separately.
  It is shown that the doping concentration derived from the capacitance-voltage measurements with and without edge corrections differ significantly.
  After the edge correction, the bulk doping of the pad diodes is found to be uniform within $\pm 1.5$\,\%.
  The voltage dependence of the edge capacitance is compared to the predictions of two simple models.

\end{abstract}

\begin{keyword}
 Silicon pad diodes \sep doping profile \sep  edge effects \sep  capacitance-voltage.
\end{keyword}

\maketitle
 \pagenumbering{arabic}

\section{Introduction}
 \label{sect:Introduction}
 Capacitance-voltage (\textit{C--V})\,measurements of diodes are a standard method to determine doping profiles.
 In pad diodes, fabricated on high-ohmic ($\gtrsim 1$\,k$\Omega$\,cm) silicon and thicknesses of several $100\,\upmu$m, the guard rings and the lateral extension of the depletion zone significantly contribute to the capacitance.
 In Ref.\,\cite{Copeland:1970} the influence of edge effects on doping-profile determinations is presented, and a parametrization for the corresponding capacitance derived.
 However, this formula does not apply to diodes with guard rings.
 Using two $p^+n$ pad diodes from the same wafer with similar guard rings but different areas, the planar capacitance per unit area, $C_{planar}$, and the edge capacitance per unit length, $C_{edge}$, have been determined and their impact on different determinations of the doping density
 investigated.
 \textit{C--V}\,measurements were made with the guard ring (GR) floating and with the GR at the same potential as the $p^+$\,implant.
 Significant differences are found.
 It is concluded that, for a precise determination of the doping from \textit{C--V}\,measurements, edge effects have to be taken into account, even if both diode and GR are at the same potential.
 It is also recommended that pad diodes with different areas are implemented as standard test structures when sensors are fabricated on high-ohmic silicon.



 \section{Pad diodes investigated and measurement setup}
  \label{sect:Sensors}

\begin{figure}[!ht]
   \centering
   \begin{subfigure}[a]{0.5\textwidth}
    \includegraphics[width=\textwidth]{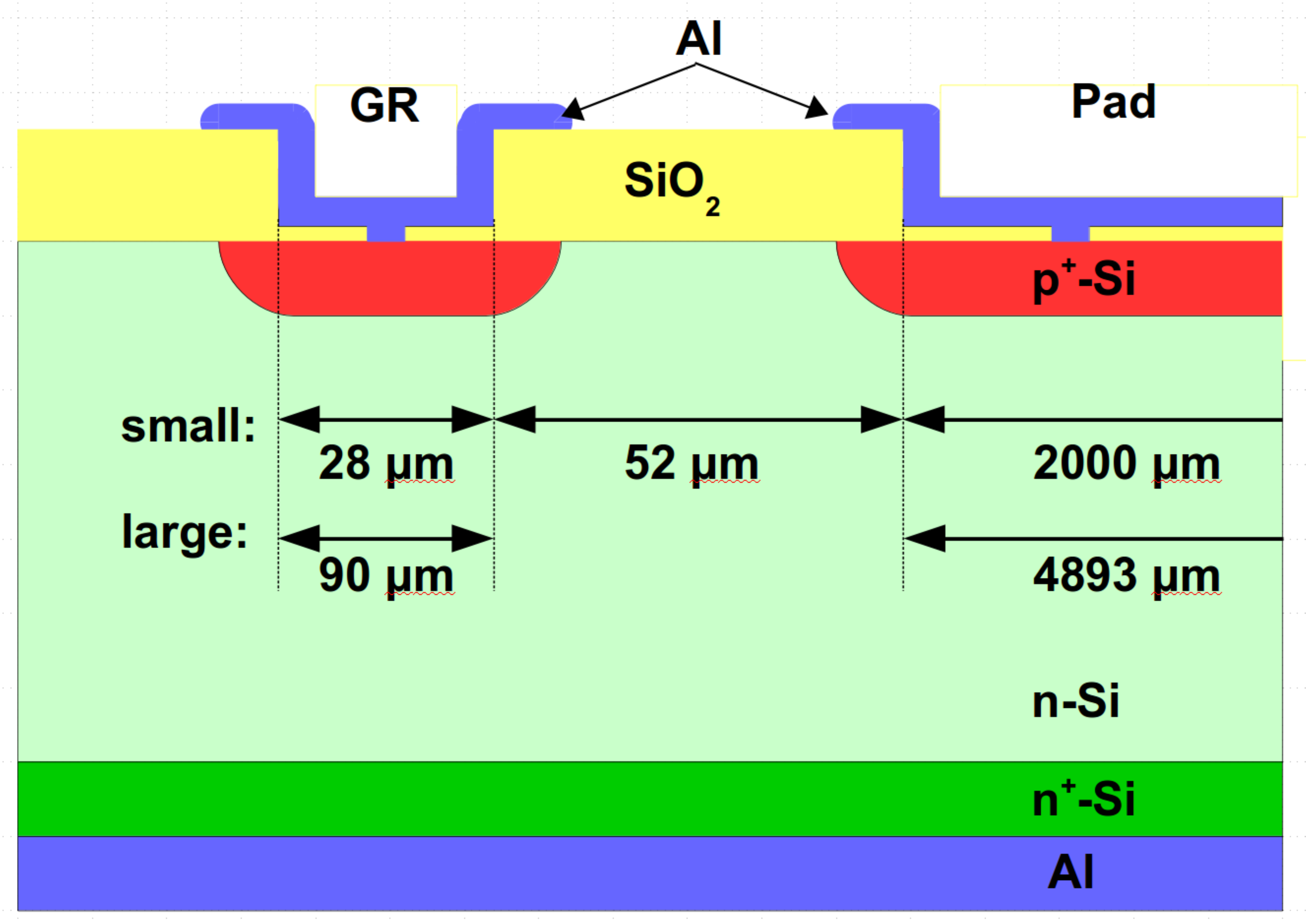}
    \caption{ }
   \end{subfigure}%
    ~
   \begin{subfigure}[a]{0.5\textwidth}
      \vspace{0.7cm}
    \includegraphics[width=\textwidth]{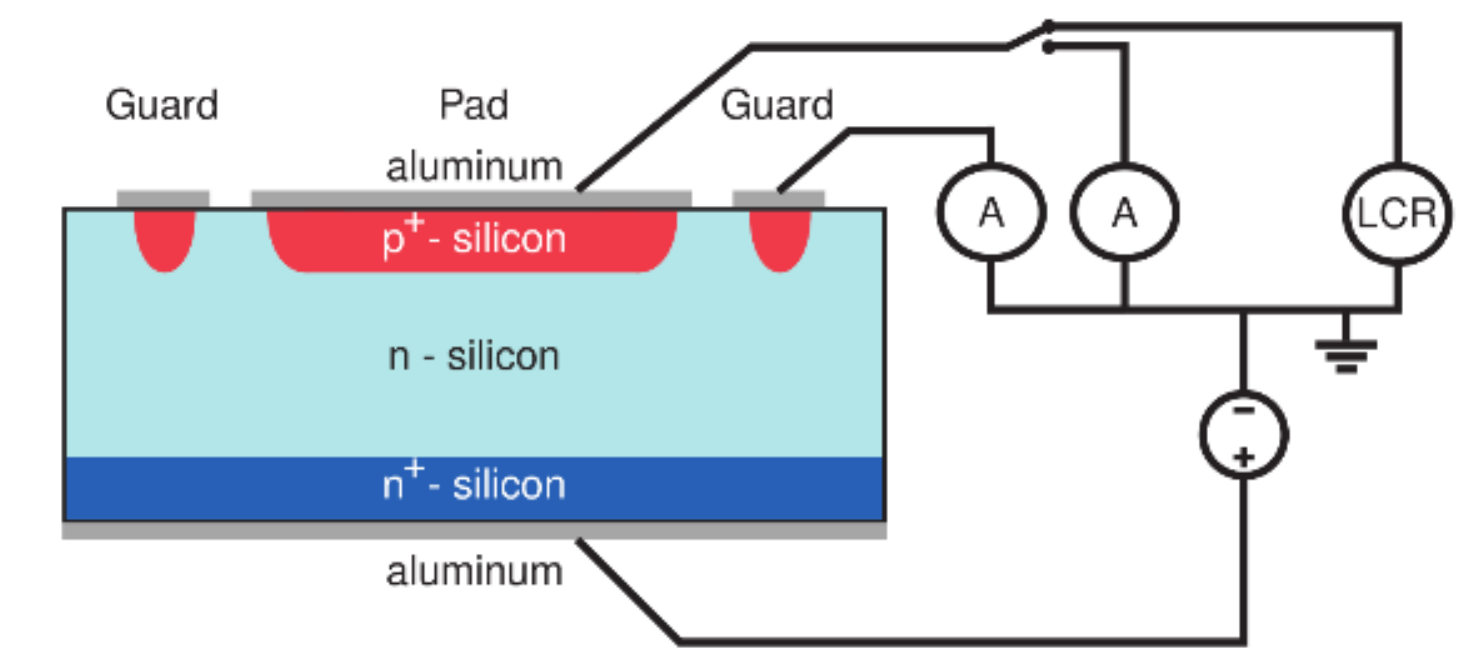}
      \vspace{0.2cm}
    \caption{ }
   \end{subfigure}%
   \caption{ (a) Schematic cross section of the pad diodes, and (b) measurement setup. }
  \label{fig:Fig1}
 \end{figure}

 Two square pad diodes, fabricated by Hamamatsu\,\cite{Hamamatsu} on phosphorous-doped float-zone silicon, were used for the studies.
 Fig.\,\ref{fig:Fig1}\,a shows a schematic cross section.
 The openings of the implantation windows for the pad diodes are $2000\,\upmu$m and $4893\,\upmu$m for the small and for the large diode, respectively; the corresponding numbers for the guard-ring implants are $28\,\upmu$m and $90\,\upmu$m.
 Guard-ring and pad-diode implants are separated by $52\,\upmu$m of SiO$_2$.
 These dimensions have been obtained from the GDS files.
 Assuming an implantation depth of $d_j = 2\,\upmu$m and a sideways extension of the doping below the SiO$_2$ of $0.8 \, d_j$\,\cite{Wolf:1986} results in effective implantation widths of $a_S = 2003 \pm 1 \,\upmu$m and $a_L = 4896 \pm 1 \,\upmu $m for the small and the large diode, respectively.
 For the overall thickness of the diodes $204 \pm 3\upmu$m has been measured using a micrometer gauge.
 Subtracting $4\,\upmu$m for the $n^+$ and $p^+$ implantations, $3.5\,\upmu$m for the 2 aluminum layers, and $0.5\,\upmu$m for the passivation, results in an estimated active thickness of $d_{act} = 196 \pm 5 \,\upmu$m.

 Fig.\,\ref{fig:Fig1}\,b shows the measurement setup.
 The measurements were performed at $20\,^\circ $C on a temperature-controlled probe station.
 First, the current-voltage characteristics for forward and reverse bias was measured in order to check the quality of the diodes and the contacts by the probe needles.
 Then, using an Agilent LCR\,meter\,\cite{Agilent}, the \textit{C--V}\,characteristics of the two pad diodes were measured for "GR grounded", and for "GR floating".
 The reverse voltage was varied between 0 and 160\,V in 0.5\,V steps, and 17 AC\,frequencies between 100\,Hz and 2\,MHz were selected.
 For frequencies between 200\,Hz and 200\,kHz the $C$\,values agreed within $\pm 1\,\%$\,\cite{Hufschmidt:2016},  and only the results of the 10\,kHz measurements will be shown.
 The $AC$\,voltage was varied between 50 and 500\,mV.
 For voltages above 4\,V the $C$\,values agreed within $\pm 1\,\%$\,\cite{Hufschmidt:2016}, and only the results for 500\,mV will be shown.


 \section{Data analysis and results}
  \label{sect:Analysis}

 \subsection{Planar and edge capacitances }
  \label{sect:Planar}

 To separate the \emph{edge} and the \emph{planar} contribution to the total capacitance of a square diode of edge length $a$, we assume
 \begin{equation}
  \label{equ:Csum}
   C_a = a^2 \, C_{planar} + 4\, a \, C_{edge},
 \end{equation}
 and obtain from the capacitances of two square diodes with edge lengths $a_S$ and $a_L$
 \begin{equation}
  \label{equ:Csep}
   C_{planar}=\frac{\frac{C_{a_S}}{a_S}-\frac{C_{a_L}}{a_L}}{a_S-a_L}
   \hspace{1cm}\textrm{and}\hspace{1cm}
   C_{edge}=\frac{ \frac{a_L} {a_S} C_{a_S} - \frac{a_S} {a_L} C_{a_L} }{4\,(a_L-a_S)}.
 \end{equation}
 It has been verified that the results change by less than 0.1\,\%, if the
 changes in perimeter and area due to the rounding at the 4 corners (radius of 79\,$\upmu $m) are taken into account in the analysis.
 Note that $C_{planar}$ refers to the capacitance per unit area, and $C_{edge}$ to the capacitance per unit length.

 \begin{figure}[!ht]
   \centering
   \begin{subfigure}[a]{0.5\textwidth}
    \includegraphics[width=\textwidth]{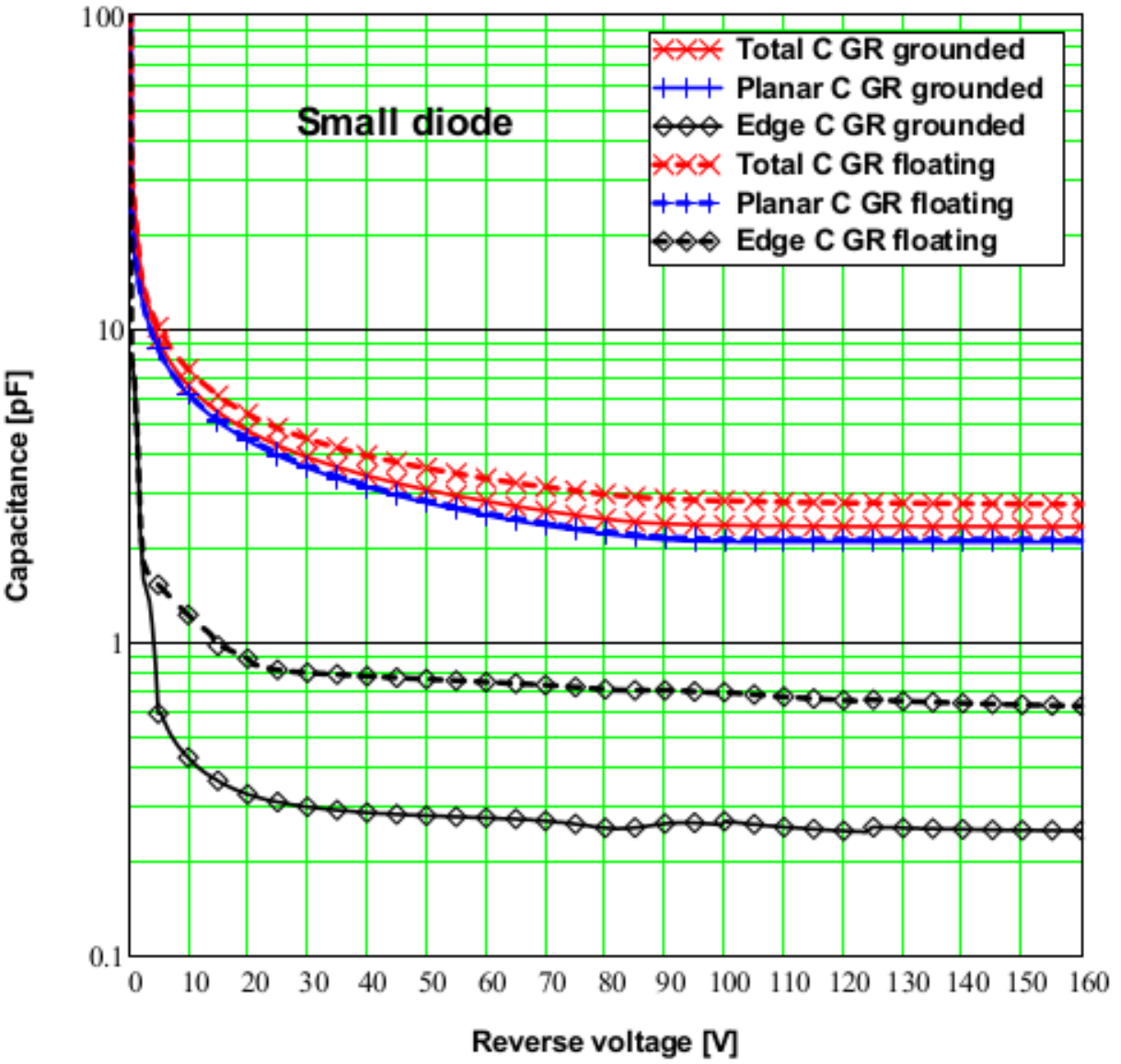}
    \caption{ }
   \end{subfigure}%
    ~
   \begin{subfigure}[a]{0.5\textwidth}
    \includegraphics[width=\textwidth]{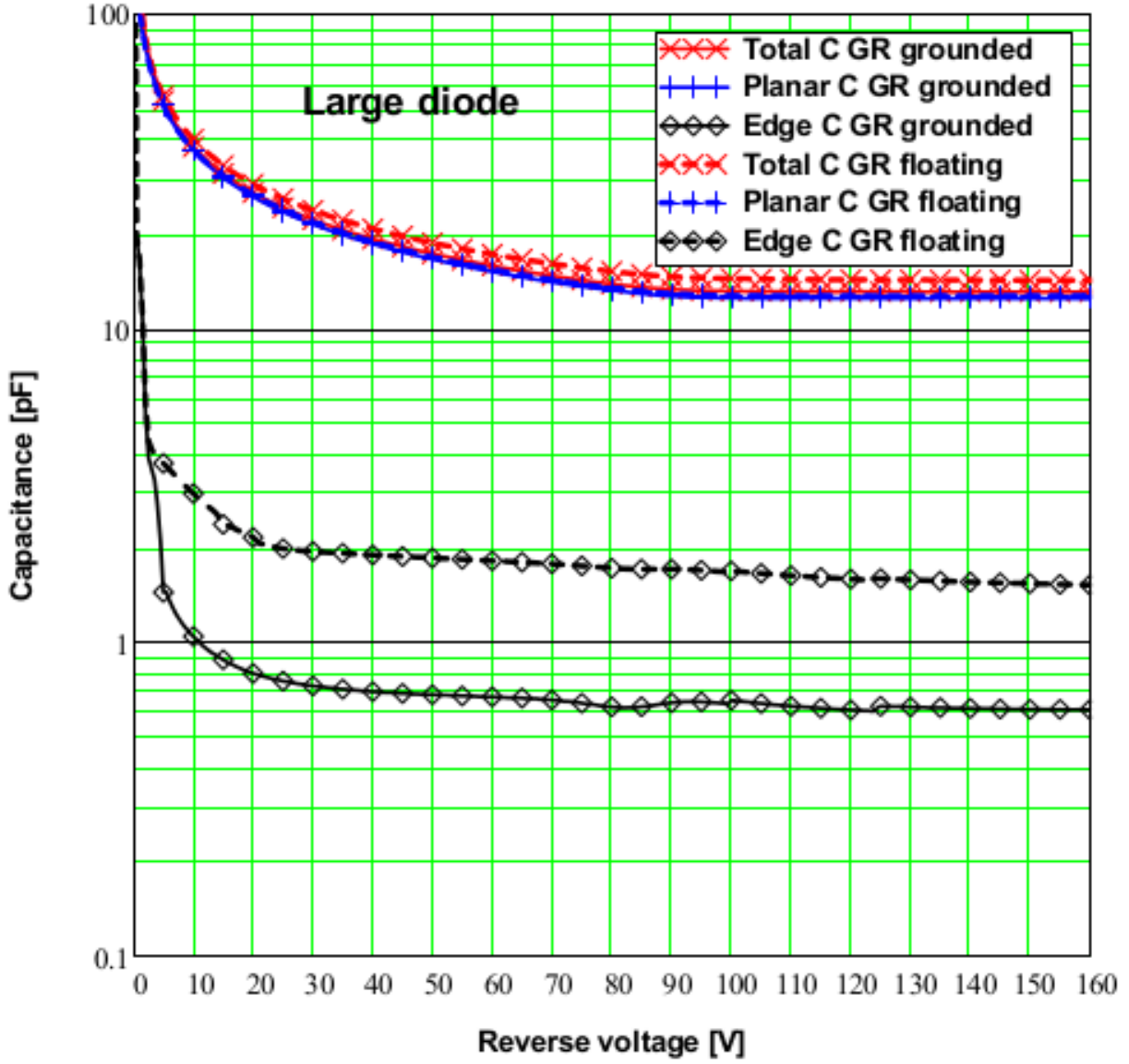}
    \caption{ }
   \end{subfigure}%
   \caption{Separation of the measured total capacitance into the planar and edge contributions using Eq.\,\ref{equ:Csep} for (a) the small diode, and (b) the large diode, with the GR grounded and the GR floating.}
  \label{fig:CV}
 \end{figure}

 Fig.\,\ref{fig:CV} shows the measured total capacitance and the separate contributions of planar and edge capacitances for the small and the large diodes, derived from the measurements with the GR grounded and the GR floating.
 Whereas, as expected, the total capacitance is larger for "GR floating" than for "GR  grounded", the planar capacitances determined are the same.
 This demonstrates that the separation of the planar and edge contributions is successful.
 As the ratio of the perimeter to the area of a square diode is proportional to 1/\textit{a}\,, the relative contribution of the edge capacitance is expected to decrease with diode side length.
 This is actually observed: The numbers for GR\,floating/GR\,grounded are $\approx 37/13$\,\% for the small, and $\approx 10/4$\,\% for the large diode.
 The grounding of the GRs reduces the edge contribution by a factor of about 3, but does not eliminate it.
 It can be noticed that the planar capacitance saturates at the depletion voltage of $\approx \,90$\,V, which is not the case for the edge capacitance with GR floating. In the latter case, the depletion region increases sideways with increasing voltage, and causes a further decrease of the capacitance.
 It is also observed that at lower voltages the voltage dependence of the edge and of the planar capacitances are quite different.
 Therefore, the doping profiles extracted with and without a correction of  edge effects will be different, as will be discussed in the next section.

 \subsection{Doping density determinations}
  \label{sect:Doping}

\begin{figure}[!ht]
   \centering
   \begin{subfigure}[a]{0.52\textwidth}
    \includegraphics[width=\textwidth]{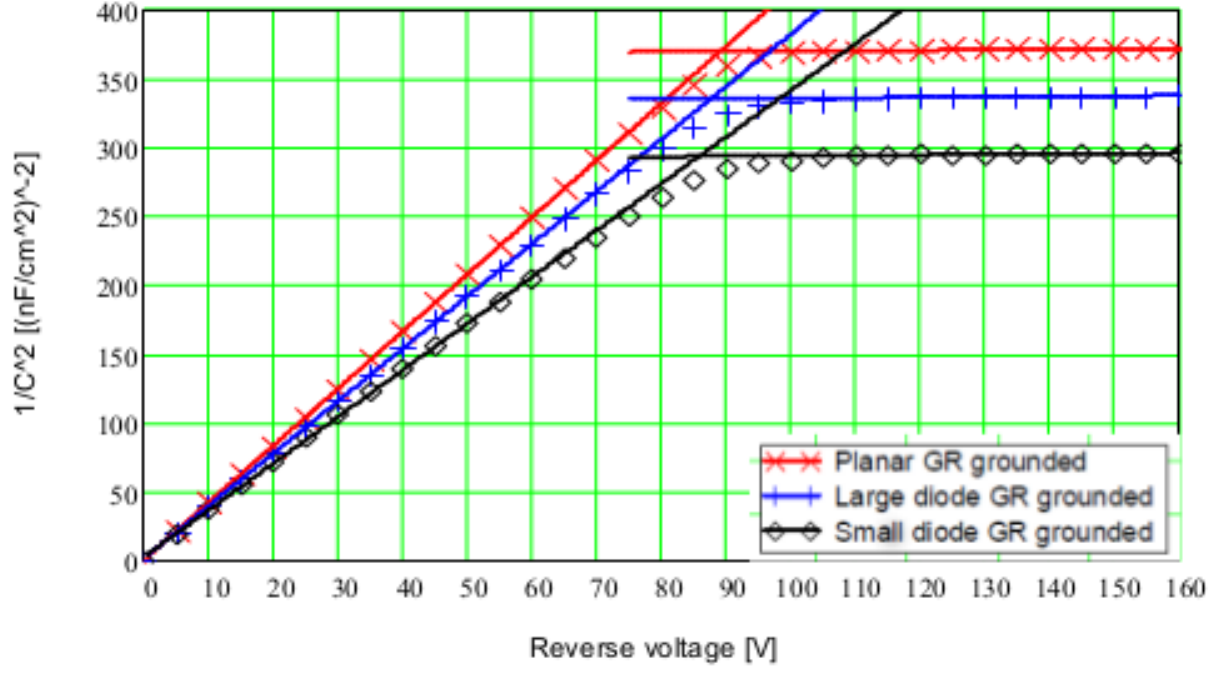}
    \caption{ }
   \end{subfigure}%
    ~
   \begin{subfigure}[a]{0.48\textwidth}
    \includegraphics[width=\textwidth]{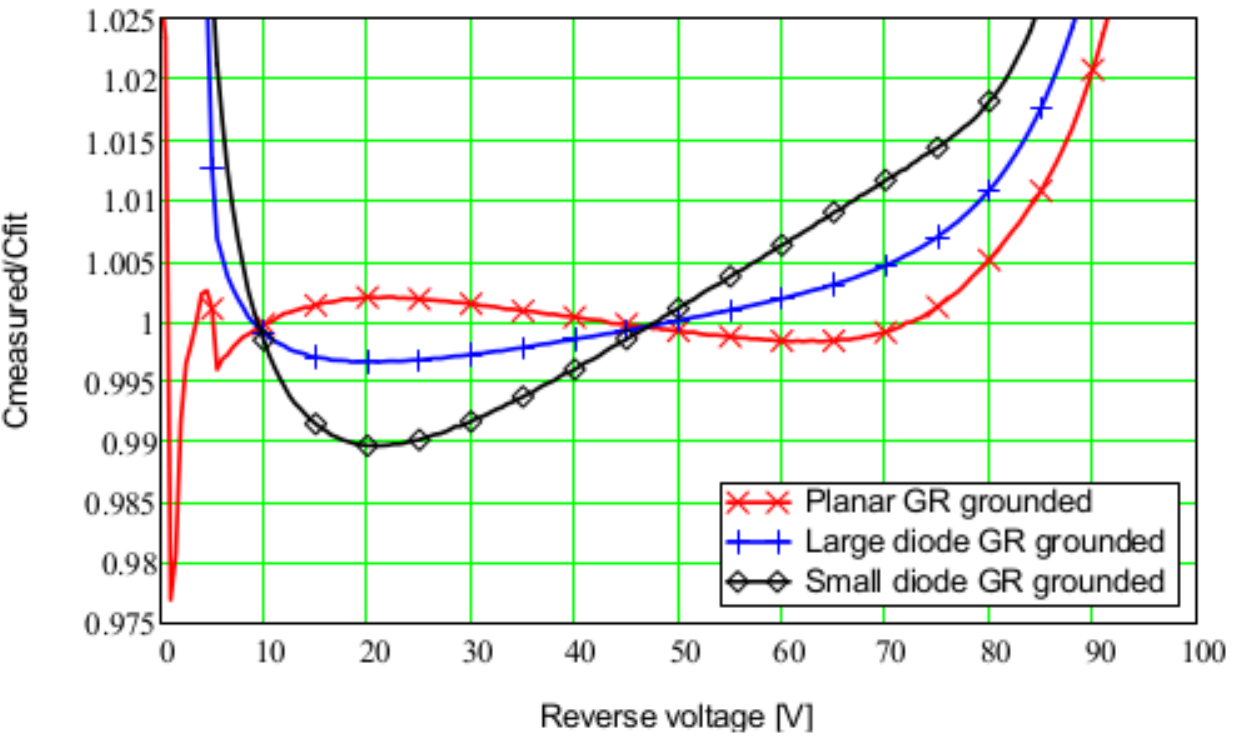}
    \caption{ }
   \end{subfigure}%
   \caption{(a) Values of $1/C^2$ normalized to 1\,cm$^2$, for $C_{planar}$, $C_{a_L}$ and $C_{a_S}$ for the capacitance measurements with the GR grounded and floating, as a function of the reverse voltage.
   The curves are the straight-line fits below and above the full-depletion voltage described in the text.
   (b) Ratio of the measured to the fitted capacitance value for $C_{planar}$, $C_{aL}$ and $C_{aS}$, with the GR grounded.
   A linear function was fitted to $1/C(V)^2$ for the reverse voltage between 5 and 75\,V.  }
  \label{fig:Fig2}
 \end{figure}

  Fig.\,\ref{fig:Fig2}\,a shows  the dependence of the inverse square of the measured capacitance per unit area as a function of the reverse voltage for $C_{planar}$, $C_{a_L}/a_L^2$ and $C_{a_S}/a_S^2$ for the measurements with GR grounded.
  We show $1/C(V)^2$, as in the approximation of an abrupt junction and uniform doping, a straight line with the slope $2/(q_0 \, \varepsilon _0 \, \varepsilon _{Si} \,N_D ) $ is expected up to the depletion voltage $V_{d}$\,\cite{Grove:1967}.
  The doping density of the $n$-type Si of the $p^+n$\,diode is denoted $N_D$, $q_0$ the elementary charge, and $\varepsilon _0 \, \varepsilon _{Si}$ the dielectric constant of Si.
  Table\,\ref{tab:Table1} presents parameters extracted from the  \textit{C--V}\,measurements.

 \begin{table} [!ht]
  \centering
   \begin{tabular}{c||c|c|c|c|c|c|c|c}
   & $a$ & $C$(120\,V) & $d$ & $V_{d}$ & $N_D^{V_{d}}$ & $N_D^{1/C^2}$ & $V_0$ & $\chi ^2$/ \\
   & [$\upmu $m] & [pF/cm$^2$] & [$\upmu $m] & [V] & [$10^{12}$/cm$^{3}$] & [$10^{12}$/cm$^{3}$] & [mV] & NDF  \\
  \hline \hline
  Planar & 10 000& 51.99 & 203 & 89.1 & 2.93 & 2.86 & 236 & 249/139 \\
    \hline
  Large diode& 4 896 & 54.54 & 193 & 87.6 & 2.88 & 3.12 & 516 & 1300/139 \\
    \hline
  Small diode& 2 003 & 58.21 & 181 & 85.7 & 2.82 & 3.50 & 913 & 9600/139 \\
  \hline  \hline
   \end{tabular}
  \caption{Parameters extracted from the \textit{C--V}\,data with the GR grounded. For explanations see text.
  \label{tab:Table1} }
 \end{table}


  The third column of Table\,1 shows $C$(120\,V), the normalized capacitances for "GR grounded", from which we derive the effective active diode thicknesses, $d$.
  As expected from edge effects, the value obtained for $d$ is smaller if the edge correction is not applied.
  The values for "planar" and the large diode, which have an estimated uncertainty of $\pm  2\,\upmu$m, are compatible with the estimated active thickness of $196 \pm  5\,\upmu$m, whereas it is significantly smaller for the small diode.
  We conclude that at least for smaller pad diodes a correction for the edge capacitance is required for a precise determination of the active diode thickness from the capacitance measured above the depletion voltage.

  For the quantitative analysis, the $1/C(V)^2$ data are fitted by linear functions $b\,(V - V_0)$ below (5 -- 75\,V) and above (110 -- 160\,V) the full depletion voltage, $V_{d}$, as shown in Fig.\,\ref{fig:Fig2}\,a.
  A measurement error of $0.1\,\%\, C$ is assumed for the fit.
  Fig.\,\ref{fig:Fig2}\,b\,shows for the fit below $V_{d}$, the ratio of measured to fitted $C$\,value as a function of the reverse voltage, and Table\,\ref{tab:Table1} gives the results.
  "Planar" denotes the results for $C_{planar}$, according to Eq.\,\ref{equ:Csum}.
  Whereas the maximum difference between fit and data below $V_{d}$ is $0.3\,\%$ for "planar", it increases to 0.7\,\% and 1.5\,\% for "GR grounded" for the large and small diode, respectively.
  From Fig.\,\ref{fig:Fig2}\,b it is obvious that the point-to-point measurement uncertainties are well below 0.1\,\%.
  Nevertheless, the agreement at the permille level after correcting for edge effects, is quite satisfactory.
  We conclude that for the pad diodes investigated, only after correcting for the edge capacitance, $1/C^2$ is to a good approximation a linear function of the reverse voltage.

  From the slope $b$ from the fits for $V < V_d$, the $n$-doping density $N_D^{1/C^2} = 2/(q_0\,\varepsilon _0 \, \varepsilon _{Si} \, b) $ is obtained.
  The results for "GR grounded" are shown in Table\,\ref{tab:Table1}.
  The values for "planar" and for the large and small diodes are quite different, showing the impact of the edge effects.

  From the intercept $V_0$ from the fits for $V < V_d$, information on the built-in voltage, $V_{bi}$, is obtained.
  Frequently $V_{bi} = |V_0| + 2\,k_\mathrm{B}T/q_0$ is assumed\,\cite{Grove:1967}.
  However, in Ref.\,\cite{Chang:1967} it is concluded:
  \textit{The barrier height cannot be determined accurately by capacitance measurements.
  The true 1/$C^2$\,intercept is less than the barrier voltage.}
  The typical corrections given in \,\cite{Chang:1967} are between 0.1 and 0.4\,V.
  We conclude that the value found for $V_0$ agrees with expectations, however it does not allow to precisely determine the built-in voltage.

  From the voltage at which the straight lines of the fits above and below depletion cross (see Fig.\ref{fig:Fig2}\,a), the full-depletion voltage, $V_d$, is obtained.
  The corresponding value for the doping density  $N_D^{V_d} \approx [(2\,\varepsilon _0 \, \varepsilon _{Si})/(q_0\,d^2)]\, V_d $, where  $d = 200\,\upmu $m is  used for the diode thickness.
  The results are  shown in Table\,\ref{tab:Table1}.
  The values for $N_D$ agree  within $\approx \pm 2$\,\%.
  However, below $V_d$ only the fit for planar is acceptable, as can be seen from the increase in $\chi ^2$ shown in Table.\,\ref{tab:Table1}.
  For "planar"  the values of $N_D^{V_d}$ and of $N_D^{1/C^2}$ agree within the estimated uncertainties.
  This however, is not the case for the large and small diodes.
  We conclude that only if the correction for edge effects is applied, the effective doping densities obtained from the measurement of the $1/C^2$\,slope and of $V_d$ are consistent.
  Without correction, the determination of the doping from $V_d$ has a smaller uncertainty than the one from the $1/C^2$\,method.

\begin{figure}[!ht]
   \centering
    \includegraphics[width=0.5\textwidth]{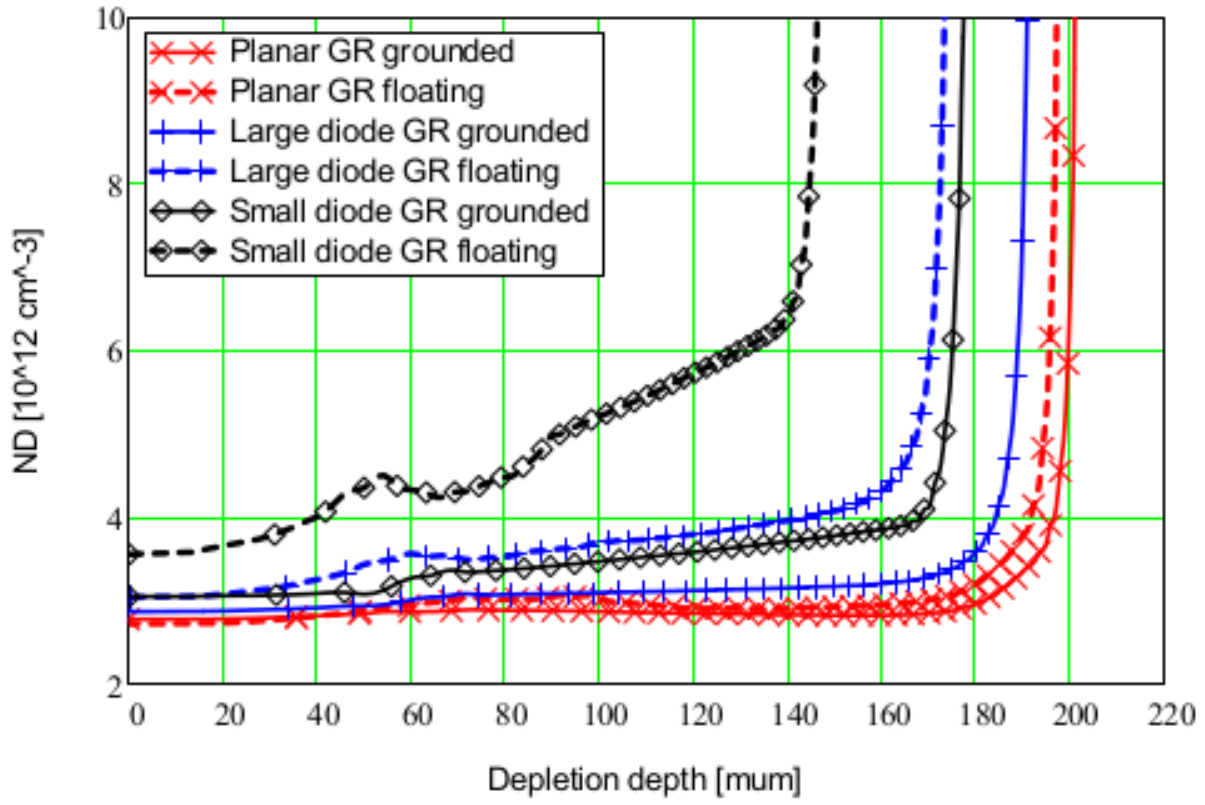}
    \caption{Doping-density profiles determined from the capacitance measurements of "planar", the large and the small diode, for the conditions "GR grounded" and "GR floating" using Eq.\,\ref{eq:NDx}. }
   \label{fig:Fig3}
 \end{figure}

  The doping density $N_D(x)$ of the $n$\,region as a function of $x$,  the distance from the $p^+n$\,junction can be determined using\,\cite{Grove:1967}
  \begin{equation}
   \label{eq:NDx}
    x(V) = \frac{\varepsilon _0 \,\varepsilon _{Si}}{C(V)} \hspace{1cm} \mathrm{and} \hspace{1cm}
    N_D(x(V)) = \frac{2}{q_0\,\varepsilon _0 \,\varepsilon _{Si}} \frac{1}{\frac{\mathrm{d}(1/C^2)}{\mathrm{d}V}}.
  \end{equation}
 The results for "planar", "GR grounded" and "GR floating" are shown in Fig.\,\ref{fig:Fig3}.


 An approximately constant $N_D(x)$ is expected up to the $x$\,value at which the effect of the backside $n^+$\,implant becomes relevant.
 This is actually observed for "planar", where $N_D(x)$ is constant within $< \pm 1.5$\,\%.
 A linear increase with $x$ by about 10\,\% is found for the large diode with the GR grounded, and significantly larger changes for the other measurements.
 We conclude that for a precise determination of the doping profile, in particular for small diodes, the edge correction has to be applied.

 In order to quantify the influence of the edge correction on the determination of the doping profile, $N_D(x)$, the measured $C(V)$ dependence is  fit to the predictions for an assumed parametrization of $N_D(x)$.
 The advantage of this approach is that the derivative $\mathrm{d} C/ \mathrm{d} V$, which usually requires smoothing and makes the proper treatment of the measurement errors difficult, is not required.
 A similar approach is discussed in Ref.\,\cite{Yaremchuk:2001}.
 From the one-dimensional Poisson equation
 \begin{equation}\label{equ:Poisson}
   N_D(x) = - \frac{\varepsilon _0 \, \varepsilon_{Si} } {q_0} \, \frac {\mathrm{d}^2 V(x) } {\mathrm{d} x^2 }
   \hspace{0.5cm} \mathrm{follows} \hspace{0.5cm}
   V(w) = \frac{q_0} {\varepsilon _0 \, \varepsilon_{Si} } \int_0^w \Big( \int_\xi^w N_D(\xi ')\, \mathrm{d} \xi '\Big) \mathrm{d} \, \xi,
 \end{equation}
 where $V(w)$ is the voltage drop over the depletion region of width $w$.
 Assuming a linear dependence of the doping density on the distance $x$ from the junction,
  \begin{equation}\label{equ:ND}
    N_D(x)=N_0+\alpha \,x,
  \end{equation}
 and the voltage $V_0$, which is related to the built-in voltage,
 \begin{equation}\label{equ:Cw}
   V(w) = \frac{q_0} {\varepsilon _0 \, \varepsilon_{Si}}\,w^2 \, \Big( \frac{N_0} {2} + \frac{\alpha \,w} {3} \Big) - V_0
   \hspace{0.25cm} \mathrm{and}  \hspace{0.25cm}
   V(C) = \frac {q_0 \, \varepsilon _0 \, \varepsilon_{Si}} {C^2} \, \Big( \frac {N_0} {2} + \frac{\alpha \,\varepsilon _0 \, \varepsilon_{Si}} {3\,C} \Big) - V_0
 \end{equation}
 is derived.
 Here the relation $  w(C) = \varepsilon _0 \, \varepsilon_{Si}/C $  is used.
 For the fit, $C(V)$ is calculated using a linear interpolation on an 1\,$\upmu $m grid of $\ln [V(w)]$ and $\ln [C(w)]$.
 We note that for $\alpha = 0$ Eq.\,\ref{equ:Cw} gives the well known relation for the abrupt single-sided junction $ C(V) \propto (V+V_0)^{-1/2} $, and for $N_0 = 0$, the linearly-graded junction, $ C(V) \propto (V+V_0)^{-1/3} $.
 These are special cases for the more general case:
 For $N_D(x) \propto x^n$ Eq.\,\ref{equ:Poisson} gives $C(V) \propto (V+V_0)^{-1/(n+2)}$.
 An equivalent relation has been derived in Ref.\,\cite{Coerver:1970}.

 \begin{figure}[!ht]
   \centering
    \includegraphics[width=0.5\textwidth]{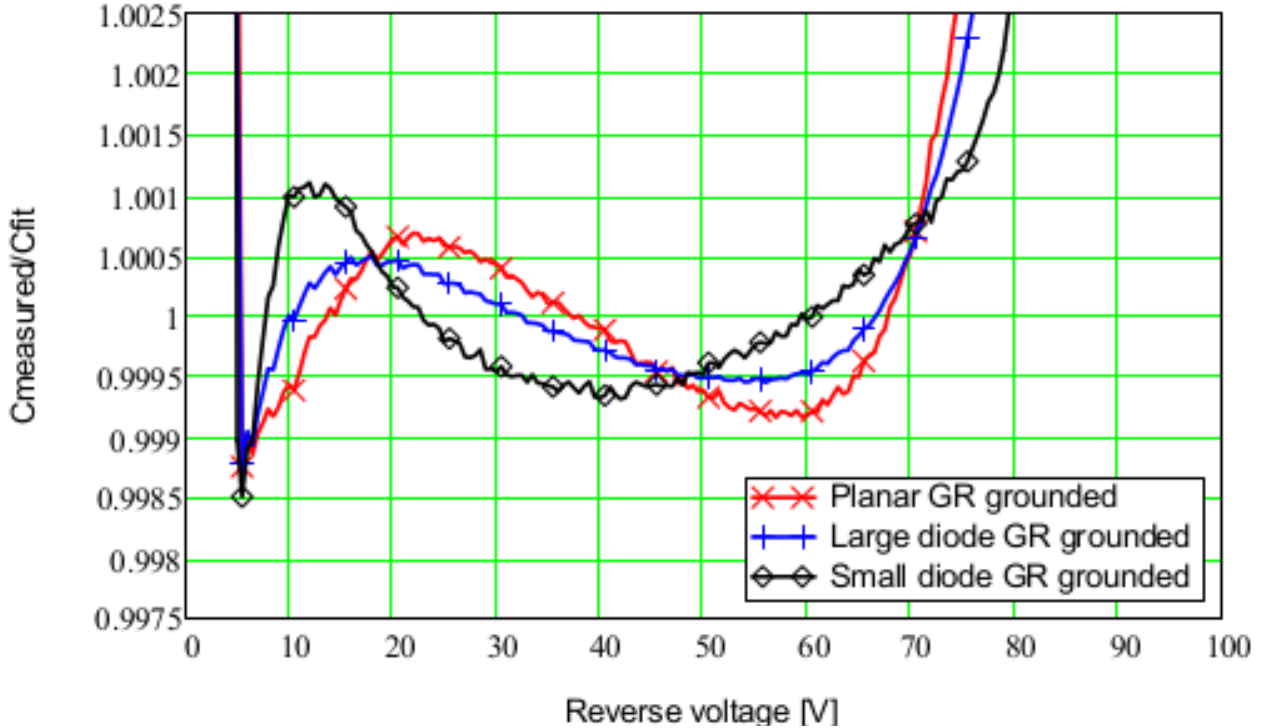}
     \caption{Ratio of the measured to the fitted capacitance value for $C_{planar}$, $C_{aL}$ and $C_{aS}$, with the GR grounded, using the doping-profile parametrisation given in Eq.\,\ref{equ:ND}.}
  \label{fig:Fig3c}
 \end{figure}

  \begin{table} [!ht]
  \centering
   \begin{tabular}{c||c|c|c|c|c|c}
   & $N_0$ & $\alpha $ & $V_0$ & $<N_D>$ & $\Delta N_D$ & $\chi ^2$/ \\
   & [$10^{12}$/cm$^{3}$] & [$10^{12}$/cm$^{4}$] & [mV] & [$10^{12}$/cm$^{3}$] & [$10^{11}$/cm$^{3}$] & NDF  \\
  \hline \hline
  Planar & 2.92& $-4.8$ & \,342 & 2.87 & $-0.48$ & 94/138  \\
    \hline
  Large diode& 2.95 & \,15.9 & \,197 & 3.10 & \,\,\,\,1.59 & 45/138 \\
    \hline
  Small diode& 2.95 & \,53.6 & $-29$ & 3.49 & \,\,\,5.36 & 51/138 \\
  \hline  \hline
   \end{tabular}
  \caption{Parameters extracted from the \textit{C--V}\,data with the GR grounded for the fit using Eq.\,\ref{equ:Cw}.
  \label{tab:Table2} }
 \end{table}

 Fig.\,\ref{fig:Fig3c} and Table\,\ref{tab:Table2} show the results of the fit for the voltage range $5 -75$\,V.
 The quality of the fits are significantly improved compared to the linear $1/C^2$\,fits, and the deviations of data minus fit are below 0.1\,\%.
 The values for the average doping densities, <$N_D$>, agree with the values from the $1/C^2$\,fits, shown in Table\,\ref{tab:Table1}.
 $\Delta N_D$ gives the deviation of the doping density at the diode surfaces and the average doping density.
 The values found are 1.7\,\%, 5.1\,\%, and  15\,\% for "planar", the small and the large diode, respectively.
 Again, the values agree with the  values from the  $1/C^2$\,fits, shown in Fig.\,\ref{fig:Fig3}.
 However, the values found for $V_0$ are quite different, in particular for the small and large diodes without the edge correction.
 Thus we reach the same conclusions as for the $1/C^2$\,fits: Edge effects have to be taken into account for a precise determination of the doping profiles of diodes built on high-ohmic silicon. 

 Using a method similar to Ref.\,\cite{DeMan:1970}, we have verified that the correction due to the voltage-dependent extension of the depletion zone into the $p^+$\,region, which is less than 0.2\,$\upmu $m, is small and can be neglected.
 We have also checked that the fact that \textit{C--V} measurements actually determine the density of majority carriers and not the doping density as discussed in Ref.\cite{Wilson:1980}, has a negligible effect on the results.

 \subsection{Edge capacitance}
  \label{sect:Edge}

\begin{figure}[!ht]
   \centering
   \begin{subfigure}[a]{0.5\textwidth}
    \includegraphics[width=\textwidth]{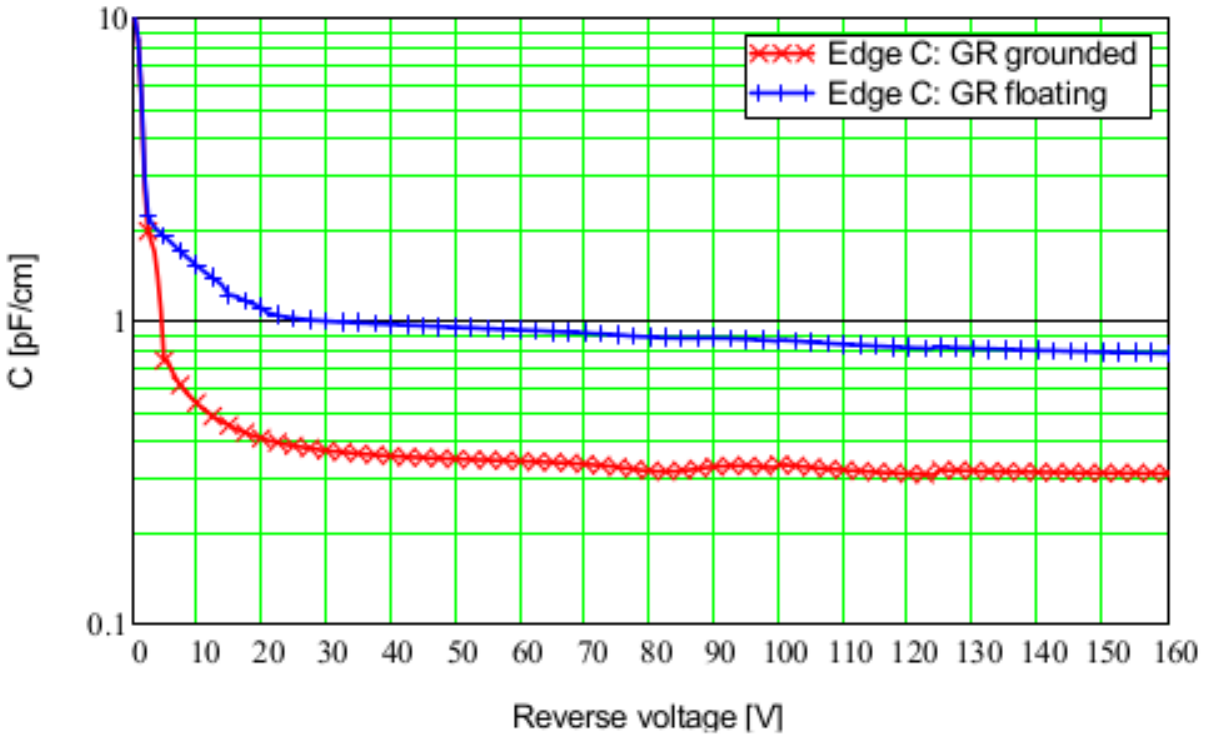}
    \caption{ }
   \end{subfigure}%
    ~
   \begin{subfigure}[a]{0.5\textwidth}
    \includegraphics[width=\textwidth]{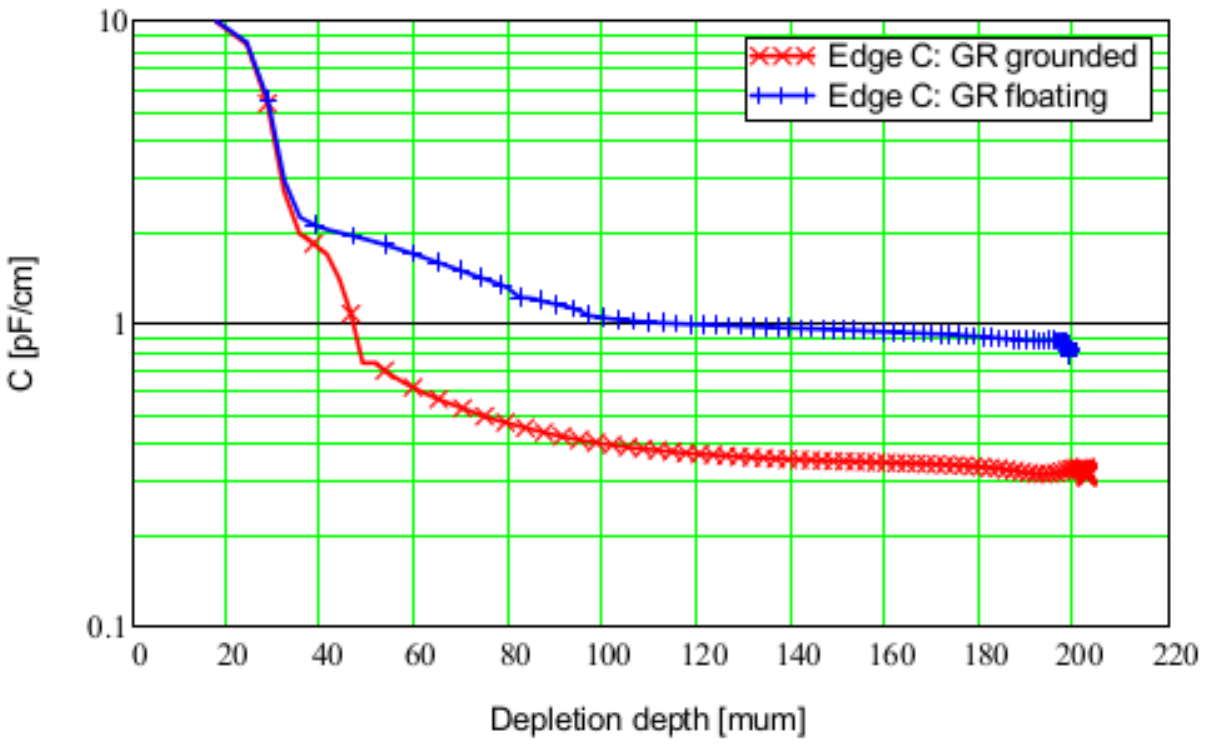}
    \caption{ }
   \end{subfigure}%
   \caption{Edge capacitance per unit length, $C_{edge}$, for "GR grounded" and "GR floating", (a) as a function of the reverse voltage, and (b) as a function of the depletion depth of the planar part of the pad diode.  }
  \label{fig:Fig4}
 \end{figure}

 The values of the edge capacitance per unit length, $C_{edge}$, determined using Eq.\,\ref{equ:Csep}, is shown in Fig.\,\ref{fig:Fig4} as a function of the reverse voltage and as a function of the depletion depth of the planar part of the pad diode.
 As expected, $C_{edge}$ is significantly smaller for "GR grounded" than for "GR floating", demonstrating the partial shielding of the edge capacitance by the grounded GR.

 From the formulae given in Ref.\,\cite{Copeland:1970}
 $C_{edge}^{circular}=(2\,\pi\,\varepsilon _0 \,\varepsilon _{Si}\,b)\cdot (Area/Circ)$
 can be obtained for the edge capacitance of a circular diode with the area $Area = r^2\,\pi $ and the circumference $Circ = 2\,r\,\pi$.
 For silicon $b \approx 1.5$ for a metal Schottky diode and $\approx 0.46$ for a cylinder geometry, which approximates a mesa-type diode.
 The Schottky diode can be considered a first approximation for the situation "GR floating", and the mesa-type diode for "GR grounded".
 Assuming that the edge capacitance of a square and a circular capacitance with the same $Area/Circ$\,\,ratio is the same, the edge capacitance per unit length of a square is $C_{edge} = \pi \,\varepsilon _0 \,\varepsilon _{Si}\,b/2$, and thus independent of voltage and diode dimensions.
 The \textit{b}\,ratio of mesa-type and Schottky-type diodes of 0.3 approximately agrees with the $C_{edge}$\,ratio for "GR grounded" to "GR floating".
 The predicted values of $C_{edge}$ are 2.5 and 0.75\,pF/cm for $b=1.5$ and 0.46, respectively, are of the observed order of magnitude.
 However, the data show a voltage dependence, which is not predicted by the calculations of Ref.\,\cite{Copeland:1970}.

   \begin{figure}[!ht]
   \centering
   \begin{subfigure}[a]{0.5\textwidth}
     \vspace{0.8cm}
    \includegraphics[width=\textwidth]{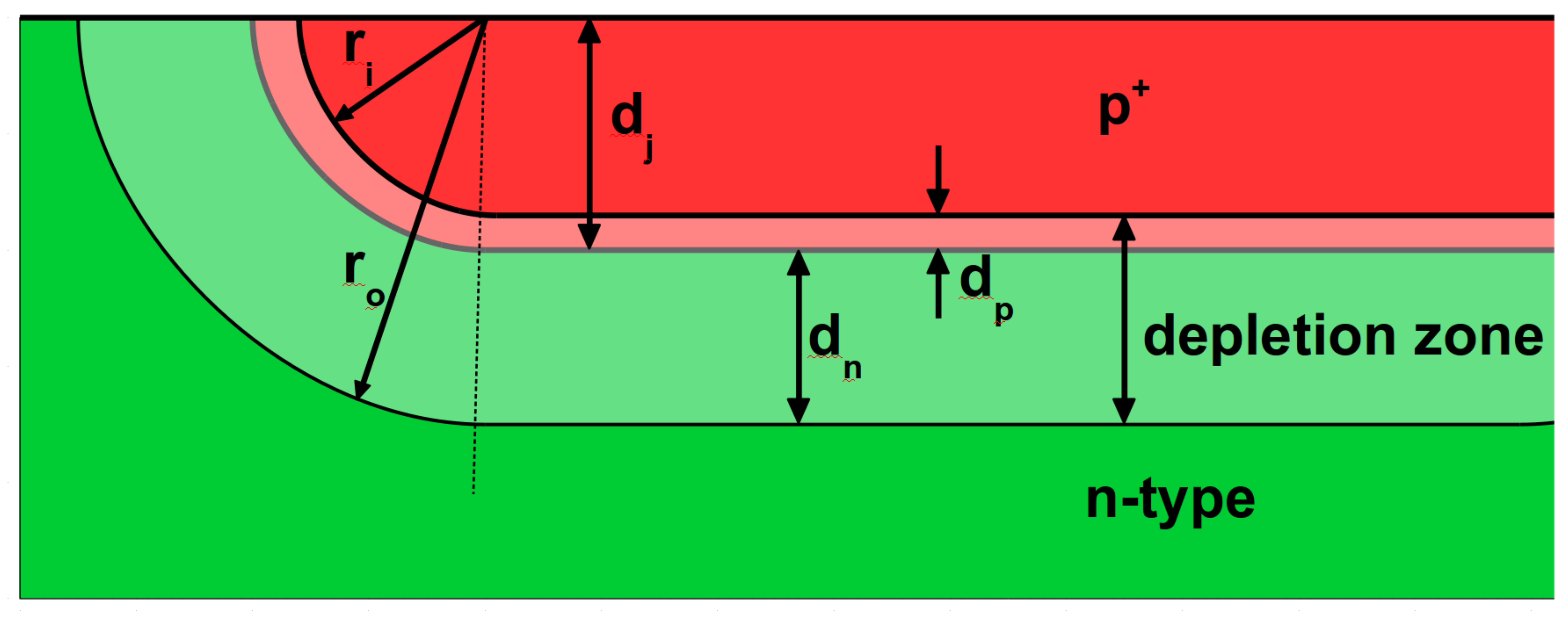}
     \vspace{0.22cm}
    \caption{ }
   \end{subfigure}%
    ~
   \begin{subfigure}[a]{0.5\textwidth}
    \includegraphics[width=0.9\textwidth]{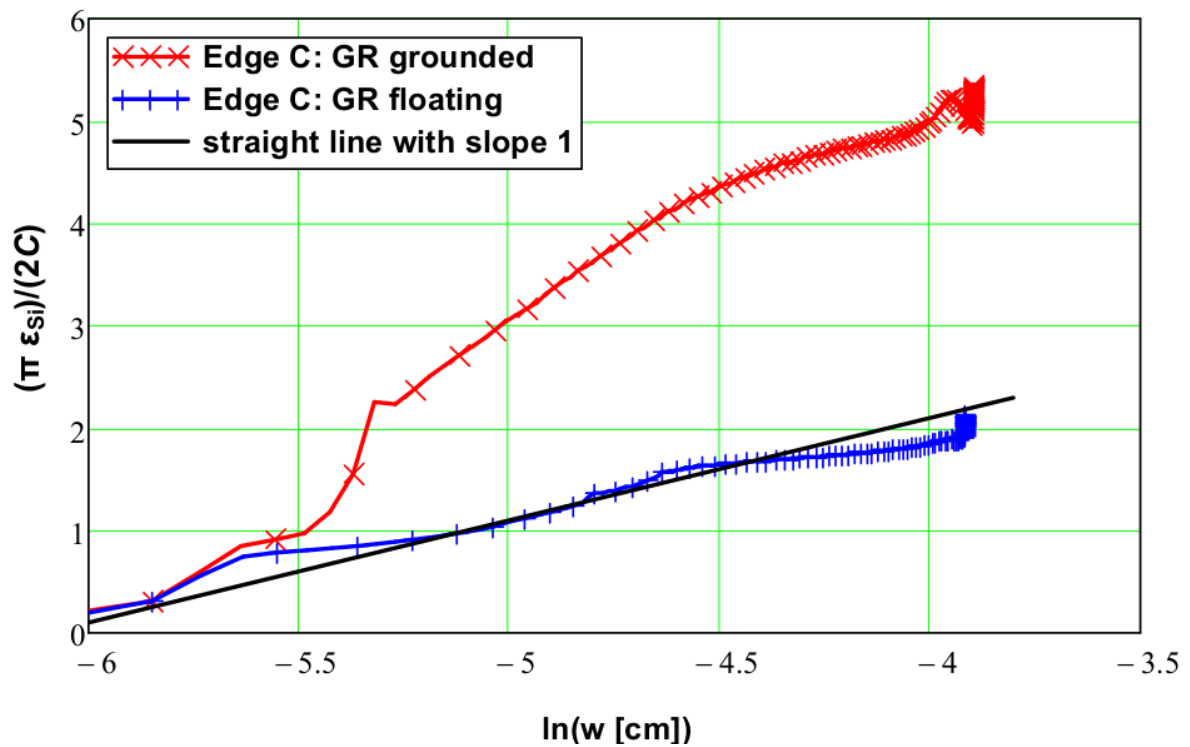}
    \caption{ }
   \end{subfigure}%
   \caption{(a) Geometrical parameters used for the model calculation of $C_{edge}$, the edge capacitance per unit length. (b) Measured value of $(\pi \,\varepsilon _0 \,\varepsilon _{Si})/(2\,C_{edge})$ as a function of $\ln (w)$, with $w$, the depletion depth of the planar capacitance of the pad diodes. The parameters of the straight line shown are discussed in the text.}
  \label{fig:Fig5}
 \end{figure}
 Another way to model the edge capacitance is illustrated in Fig.\,\ref{fig:Fig5}\,a. The capacitance of a quarter cylinder capacitor with inner radius $r_i = d_j - d_p$ and outer radius $r_o = d_j + d_n$ is:
 \begin{equation}\label{Cedge}
  C_{edge} = \frac{2\,\pi\,\varepsilon _0 \,\varepsilon _{Si}}{\ln(r_o/r_i)}\cdot \frac{1}{4}
  \approx \frac{\pi\,\varepsilon _0 \,\varepsilon _{Si}}{2}\cdot \frac{1}{\ln(w/d_j)}.
 \end{equation}
 The $p^+$\,junction depth is denoted $d_j$, the depletion depths in the $p^+$ and $n$\,\,regions $d_p$ and $d_n$, respectively, and the total depletion depth $w = d_p + d_n$.
 The approximation assumes $w \gg d_j \gg d_p$, which is well satisfied for reverse voltages above a few volts.
 From Eq.\,\ref{Cedge} follows
  $(\pi\,\varepsilon _0 \,\varepsilon _{Si})/(2\, C_{edge }) = \ln{w} - \ln{d_j}$:
 A straight line as function of $\ln{w}$ with the slope 1 and the intercept $\ln{d_j}$.
 Fig.\,\ref{fig:Fig5}\,b\, shows that this relation approximately describes the "GR floating" data.
 The straight line shown has the slope 1 and the intercept $-6.1$, corresponding to $d_j = 22\,\upmu $m.
 However, the value expected from the technology is about $2\,\upmu $m.
 Nevertheless, given the crudeness of the model, the description of the data can be considered satisfactory.

 \section{Summary and conclusions}
  \label{sect:Summary}

 In this paper, the edge effects for square $p^+n$ pad sensors with guard rings (GR) fabricated on high-ohmic silicon, are investigated.
 From the capacitance-voltage (\textit{C--V}) measurements of two diodes with different areas, the planar and the edge contributions to the total diode capacitance for the conditions "GR floating" and "GR at the $p^+$\,potential" are derived.
 Different methods are used to determine the doping density of the $n$-type region from the \textit{C--V}\,data and to study the effects of the edge corrections.
 It is found that, even for the situation with the GR at the $p^+$\,potential, the influence of edge effects is significant and has to be corrected for a precise determination of doping densities.
 Ignoring the edge corrections causes significant changes of the values and of the position dependence of the doping profiles determined.
 This is in particular true for small-area pad diodes.
 Therefore, it is recommended that pad diodes with different areas and identical GR layouts, are implemented as test structures, when sensors on high-ohmic silicon are fabricated and the bulk doping has to be known precisely.

 It is found that the ratio of the edge capacitances for "GR floating" to "GR at the $p^+$\,potential" is about 3, which shows that, even for the latter case, the GR contribution to the measured capacitance is only partially shielded.
 The dependence of the edge capacitances as a function of voltage is  compared to two models.
 A qualitative agreement is found with a model, which assumes the field of a cylinder capacitor at the diode edges.


\section*{Acknowledgements}
 \label{sect:Acknowledgement}

 We thank Peter Buhmann and Michael Matysek for maintaining the measurement infrastructure of the Hamburg Detector Laboratory, where the measurements were performed in  excellent shape, which is a necessary condition for the precision results presented in this paper. We thank Georg Steinbr\"uck and Matteo Centis Vignali for their comments.

\section{List of References}

  \label{sect:Bibliography}



\end{document}